\documentclass[12pt]{article}

\usepackage{etex}
\usepackage{graphicx}
\usepackage{dcolumn}
\usepackage{bm}
\usepackage{amssymb,amsthm,amscd,amsbsy,array}
\usepackage{amsmath,dsfont}
\usepackage{soul} 
\usepackage{graphics,xcolor}

\usepackage[all]{xy}

\numberwithin{equation}{section}

\newcommand{\half}{{\frac{1}{2}}}
\def\2{{\frac{1}{2}}}
\newcommand{\const}{\mathop{\rm const}\nolimits}

\def\bp{{\bm{p}}}

\def\bu{{\bm{u}}}

\def\br{{\bm{r}}}

\def\bg{{\bm{g}}}

\def\bp{{\bm{p}}}

\def\bx{{\bm{x}}}

\def\beq{\begin{equation}}
\def\eeq{\end{equation}}
\def\beqa{\begin{eqnarray}}
\def\eeqa{\end{eqnarray}}

\def\barray{\left(\begin{array}}
\def\earray{\end{array}\right)}
\def\barraynb{\begin{array}}
\def\earraynb{\end{array}}

\def\smallover#1/#2{\hbox{$\textstyle\frac{#1}{#2}$}} %

\newcommand{\bbR}{\mathbb{R}}
\newcommand{\la}{{\langle}}
\newcommand{\ra}{{\rangle}}
\newcommand{\fg}{{\mathfrak{{g}}}}

\newcommand{\cE}{{\mathcal{E}}}

\newcommand{\rg}{\mathrm{g}}

\newcommand{\rE}{\mathrm{E}}

\newcommand{\Ad}{{\mathrm{Ad}}}

\newcommand{\medbox}[1]{\fbox{%
\rule[-10pt]{0pt}{25pt}$\;\;\displaystyle{#1}\;\;$}%
}

\usepackage{color}
\usepackage[colorlinks=true, pdfstartview=FitV, linkcolor=blue, citecolor=blue, urlcolor=blue]{hyperref}

\begin{document}

\title{A recollection of Souriau's derivation of the Weyl equation 
via geometric quantization
}

\author{
C. Duval\footnote{mailto: duval-at-cpt.univ-mrs.fr}\\
{\normalsize Centre de Physique Th\'eorique,}\\
{\normalsize Aix Marseille Universit\'e \& Universit\'e de Toulon \& CNRS UMR 7332,}\\
{\normalsize Case 907, 13288 Marseille, France.}
}

\date{February 2, 2016}

\maketitle

\thispagestyle{empty}

\maketitle

\begin{abstract}
These notes merely intend to memorialize Souriau's overlooked achievements regarding geo\-metric quantization of Poincar\'e-elementary symplectic systems. Restricting  attention to his model of massless, spin-$\half$, particles, we faithfully rephrase and expound here Sections (18.82)--(18.96) \& (19.122)--(19.134) of his book \cite{SSD} edited in 1969. Missing details about the use of a preferred Poincar\'e-invariant polarizer are provided for completeness.
\end{abstract}


\newcommand{\SL}{\mathrm{SL}}
\newcommand{\bbC}{\mathbb{C}}
\newcommand{\Spin}{\mathrm{Spin}}
\newcommand{\bzeta}{\boldsymbol{\zeta}}
\newcommand{\boldeta}{\boldsymbol{\eta}}
\newcommand{\barzeta}{\overline{\zeta}}
\newcommand{\barZ}{\overline{Z}}
\newcommand{\diag}{\mathrm{diag}}
\newcommand{\rO}{\mathrm{O}}
\newcommand{\rU}{\mathrm{U}}
\newcommand{\cC}{\mathcal{C}}
\newcommand{\bsigma}{\boldsymbol{\sigma}}
\newcommand{\surf}{\mathsf{surf}}
\newcommand{\bxi}{\boldsymbol{\xi}}
\newcommand{\barxi}{\overline{\bxi}}
\newcommand{\bbZ}{\mathbb{Z}}
\newcommand{\rank}{\mathrm{rank}}
\newcommand{\re}{\mathrm{e}}
\newcommand{\tpsi}{\widetilde{\psi}}
\newcommand{\sfD}{\mathsf{D}}

\section{The classical model}


\subsection{Minkowski spinors: a quick review}

We start with some prerequisite. Denote by $W^{10}=\SL(2,\bbC)\ltimes\bbR^{3,1}$ the universal covering of the restricted Poincar\'e group, i.e., the neutral component of the group, $\rE(3,1)$, of isometries of Minkowski spacetime, $\bbR^{3,1}=(\bbR^4,\rg)$, where $\rg=-\vert{}d\br\vert^2+dt^2$ (we put~$c=1$).

Let us recall how the well-known spin group, $\Spin(3,1)\cong\SL(2,\bbC)$, of~$\bbR^{3,1}$ arises. 
To that end, we choose the following representation of the \textbf{Dirac matrices} acting on the Minkowskian spinor space, $\bbC^{2,2}$, namely
\begin{equation}
\gamma_j
=
\left(
\begin{array}{cc}
0&\sigma_j\\
-\sigma_j&0
\end{array}
\right)
,
\qquad
\gamma_4
=
\left(
\begin{array}{cc}
0&1\\
1&0
\end{array}
\right),
\qquad
\gamma_5
=
\left(
\begin{array}{cc}
-i&0\\
0&i
\end{array}
\right)
\label{gammas}
\end{equation}
where the $\sigma_j$, with $j=1,2,3$, denote the Pauli matrices. 

\goodbreak

All matrix representations use a tacitly chosen spinor frame with Gram matrix
$$
G
=
\half\left(
\begin{array}{cc}
0&1\\
1&0
\end{array}
\right).
$$
The Hermitian conjugate of $\zeta\in\bbC^{2,2}$ reads therefore $\barzeta=C(\zeta)^T G$ where $C:z\mapsto\overline{z}$ stands for complex conjugation, and $T$ for transposition.
The Dirac matrices are traceless and Hermitian-symmetric, i.e., $\overline{\gamma_\mu}=\gamma_\mu$ for all $\mu=1,\ldots,4$, with respect to $G$.\footnote{We have $\overline{\gamma_\mu}=G^{-1}C(\gamma_\mu)^TG$.}
They furthermore satisfy the Clifford relations
$$
\gamma_\mu\gamma_\nu+\gamma_\nu\gamma_\mu=2\rg_{\mu\nu}
$$
for all $\mu,\nu=1,\ldots4$; also 
$
\gamma_5=\gamma_1\gamma_2\gamma_3\gamma_4
$
is the \textbf{chirality} operator which anticommutes with all $\gamma_\mu$. 

\medskip$\bullet$
The \textbf{spin group} in $3+1$ spacetime-dimension may be introduced as the group of quaternionic $G$-unitary matrices commuting with the chirality. Thus $A\in\Spin(3,1)$ iff $A\in\rU(2,2)$ and $SH=HS$ where the quaternionic structure of $\bbC^4$ is defined by
$$
H
=
\left(
\begin{array}{cc}
0&JC\\
JC&0
\end{array}
\right)
$$
as well as by $A\gamma_5=\gamma_5A$; here $J=-i\sigma_2$ denotes the standard complex structure of $\bbR^2\cong\bbC$. As a consequence, we have $A\gamma(\delta{R})A^{-1}=\gamma(\varrho(A)\delta{R})$ where $\varrho(A)\in\rO(3,1)_0$.\footnote{We use the shorthand notations $\gamma(\delta{R})=\gamma_\mu\delta{R}^\mu$ for all $\delta{R}\in\bbR^{3,1}$.}
 The homo\-morphism $\varrho:\Spin(3,1)\to\rO(3,1)_0$ has kernel $\bbZ_2$. In the above representation (\ref{gammas}), the spin group is therefore generated by the matrices \cite{SSD} 
$$
A
=
\left(
\begin{array}{cc}
a&0\\
0&\overline{a^{-1}}
\end{array}
\right)\in\Spin(3,1)
\qquad
\hbox{where}
\qquad
a\in\SL(2,\bbC).
$$
where the bar denotes now Hermitian conjugation of $2\times2$ complex matrices.
\goodbreak

Introducing the two supplementary projectors
$$
\Pi=\frac{1+i\gamma_5}{2}
\qquad
\&
\qquad
\overline{\Pi}=\frac{1-i\gamma_5}{2}
$$ 
we may alternatively view the spin group as the group of those matrices
$
A=(A_1\,A_2\,A_3\,A_4)\in\Spin(3,1)
$
with column vectors \cite{Duv1} 
$$
A_1=\Pi\zeta, 
\qquad
A_2=\overline{\Pi}H\zeta, 
\qquad
A_3=\overline{\Pi}\zeta, 
\qquad
A_4=\Pi{}H\zeta
$$
where the
$\zeta\in\bbC^{2,2}$
satisfy the fundamental relations \cite{SSD}
\begin{equation}
\medbox{
\barzeta\zeta=1
\qquad
\&
\qquad
\barzeta\gamma_5\zeta=0.
}
\label{FundRels}
\end{equation}
Clearly, $\Spin(3,1)$ is parametrized by the above-mentioned spinors $\zeta$; it is thus diffeomorphic to the manifold 
$$
\Sigma^6\cong{}S^3\times\bbR^3
$$ 
of those $\zeta\in\bbC^{2,2}$ satis\-fying (\ref{FundRels}).\footnote{The diffeomorphism is thus $\Spin(3,1)\to\Sigma^6:S\mapsto\zeta=A_1+A_3$.} 

\goodbreak

If we write
\begin{equation}
\zeta=
\left(
\begin{array}{c}
\zeta'\\
\zeta''
\end{array}
\right)\in\Sigma^6
\label{zeta}
\end{equation}
with $\zeta',\zeta''\in\bbC^2\setminus\{0\}$ as half-spinors, then (\ref{FundRels}) translates as 
$
\overline{\zeta'}\zeta''=1
$
(ordinary Hermitian scalar product of $\bbC^2$).

\goodbreak

\medskip$\bullet$ The universal covering of the neutral \textbf{Poincar\'e group} is thus diffeo\-morphic to $W^{10}=\Sigma^6\times{}E^4$ described by the pairs $(\zeta,R)$ with $\zeta$ as in (\ref{FundRels}), and~$R\in{}E^4=\bbR^{3,1}$, a spacetime~event.

\medskip $\bullet$ 
With this preparation, we verify that the Minkowski vector~$P$ defined for all $\delta{R}\in\bbR^{3,1}$ by the scalar product
\begin{equation}
\medbox{
P\cdot\delta{R}=\barzeta\gamma(\delta{R})\Pi\zeta\sqrt{2}
}
\label{P}
\end{equation}
is  \textbf{null and future-pointing}.\footnote{Since $P\cdot\delta{R}=P_\mu\delta{R}^\mu$ with $P_\mu=\rg_{\mu\nu}P^\nu$, Equation (\ref{P}) may alter\-natively be read as
$P_\mu=\barzeta\gamma_\mu\left[1+i\gamma_5\right]\zeta/\sqrt{2}$,
for all $\mu=1,\ldots,4$.
} 

\goodbreak

Indeed, easy computation shows that $P\cdot\delta{R}=\overline{\zeta'}(-\sigma(\delta\br)+\delta{t})\zeta'/\sqrt{2}$, with the help of (\ref{gammas}). We then get $P\cdot\delta{R}=\cE(-\bu\cdot\delta\br+\delta{t})$ where $\bu\in{}S^2\subset\bbR^3$, and $\cE=\Vert\zeta'\Vert^2/\sqrt{2}$ with $\Vert\cdot\Vert$ the Hermitian norm of $\bbC^2$. Thus, the vector $P=\cE(\bu,1)$
is clearly as announced.

\goodbreak

Likewise, the vector~$Q$ defined by $Q\cdot\delta{R}=-\barzeta\gamma(\delta{R})\overline{\Pi}\zeta\sqrt{2}$ is null and verifies $P\cdot{}Q=-1$.

We have a projection $\pi_{WV}:W^{10}\to{}V^9$ defined by $\pi_{WV}(\zeta,R)=(P,Q,R)$. The manifold $V^9$ has already been interpreted as the \textbf{evolution space} of the model \cite{SSD,DHchiral}, and $P$ as the momentum of our particle with positive energy. We note, in view of the previous definitions of $P$ and $Q$ that $V^9=W^{10}/\rU(1)$ where the action of $\rU(1)$ is given by $e^{i\theta}(\zeta,R)=(e^{i\theta}\zeta,R)$.

\subsection{Classical motions of chiral particles}

Having in mind to describe spinning particles of half-integral spin, we find it  convenient to use the \textbf{spinor representation} of the Poincar\'e group to start with, even at the classical level. To describe the \textbf{classical dynamics} of relativistic massless particles with spin $s=\half\chi\hbar$ (where $\chi=\pm1$ is the helicity), we will hence introduce a certain $1$-form~$\varpi$ of the Lie group $W^{10}$ designed to give rise to the space of (free) motions of such particles --- as well as its prequantization (see Section \ref{prequantization}).

\medskip$\bullet$
We recall that, given a Lie group $G$ with Lie algebra $\fg$, the \textbf{canonical symplectic structure} of the coadjoint orbit, $X=G/G_{\mu_0}$, passing through $\mu_0\in\fg^*$ is constructed as follows \cite{Duv2}.\footnote{The local notation $G$ should not be confused with that of the above spinorial metric!} 
If $\theta$ is the left-invariant Maurer-Cartan $1$-form of $G$, the exterior derivative, $\sigma=d\varpi$, of the $1$-form 
\begin{equation}
\varpi=\la\mu_0,\theta\ra
\label{defvarpi}
\end{equation}
descends as the canonical symplectic $2$-form, $\omega$, of $X$, namely
$$
\sigma=\pi_{GX}^*\,\omega.
$$

\goodbreak

\medskip$\bullet$
Here, the group $G=W^{10}$ is, as above-mentioned, the direct product of $\Sigma^6$ and  spacetime $E^4$. The leaves of the characteristic distribution of $\sigma$ constitute the \textbf{space of motions} of the particle, namely $X=W^{10}/\ker\sigma$.
As to the spacetime projections of these leaves via $\pi_{WE}:W^{10}\to{}E^4$, they are interpreted as the \textbf{world-sheets} of the particle. 

\goodbreak

Denote by $\re(3,1)$ the Lie algebra of the Poincar\'e group, and choose now the origin $\mu_0=(M_0,P_0)\in\re(3,1)^*$ of~$X$ as \cite{SSD,Duv1}
$$
M_0=\chi\hbar\gamma_1\gamma_2
\qquad
\&
\qquad
\gamma(P_0)=\gamma_3+\gamma_4.
$$
Tedious calculation yields the $1$-form (\ref{defvarpi}), namely\footnote{One can write, equivalently,
$
\varpi=-P_\mu{}dR^\mu-i\chi\hbar\,\overline{\zeta}_a d\zeta^a
$
where $\overline{\zeta}_a=G_{ab}\overline{\zeta^b}$ for all spinor indices $a=1,\ldots,4$.
}
\begin{equation}
\medbox{
\varpi=-P\cdot{}dR+\chi\hbar\frac{\barzeta{}d\zeta}{i}
\label{varpi}
}
\end{equation}
where $P$ is as in (\ref{P}).

\goodbreak

We then find that $\ker\sigma$ is $4$-dimensional, that is
\begin{equation}
\delta(\zeta,R)\in\ker\sigma
\qquad
\iff
\qquad
\left\{
\begin{array}{l}
\delta\zeta
=
\displaystyle
\frac{i\chi}{\hbar}\gamma(\delta{}R)\Pi\zeta+i\lambda\zeta\\[8pt]
\delta{}R\in{}P^\perp
\end{array}
\right.
\label{kersigma}
\end{equation}
with $\lambda\in\bbR$. 

This implies, via (\ref{P}), that $\delta{P}=0$, and shows that the motions take place on null affine hyperplanes $P^\perp$ in~$\bbR^{3,1}$; the particle having a constant momentum, $P$, it is strikingly \textbf{delocalized} on the world-sheets $P^\perp$ \cite{SSD,DHchiral,DEHZ}.\footnote{The projection to the evolution space $V^9$ of the leaves of $\ker\sigma$ are the Wigner-Souriau translations of Ref. \cite{DHchiral}; see Footnote \ref{Y7}.}

\medskip$\bullet$
It hence turns out that the space of motions, $X$, is $6$-dimensional, viz.,
\begin{equation}
X^6\cong{}T^*\cC^+
\label{X6}
\end{equation}
where $\cC^+\cong\bbR^3\setminus\{0\}$ is the punctured future light-cone described by the momentum~$P$. 

\goodbreak

What about the explicit form of symplectic $2$-form, $\omega$, of $X^6$? 
Its expression is computed in a Lorentz frame where $P=(\bp,\vert\bp\vert)\neq0$, and $R=(\br,t)$. The \textit{would-be} $3$-position (spanning the fibres of $X^6\to\cC^+$) is then $\bx=\bg/\vert\bp\vert$ where $\bg$ is the boost-momentum \cite{DHchiral}. The sought expression is \cite{SSD}
\begin{equation}
\medbox{
\omega=d\bp\wedge{}d\bx-s\,\surf
\qquad
\&
\qquad
s=\half\chi\hbar
}
\label{omega}
\end{equation}
where $\surf$ is the surface $2$-form of $S^2$ described by the direction $\bu=\bp/\vert\bp\vert$.\footnote{The coordinates $\bx=(x^1,x^2,x^3)$ do  not Poisson-commute; they are sometimes called the \textbf{Pryce coordinates}.}

\goodbreak

The projection $\pi_{WX}:W^{10}\to{}X^6$ is given by
$\pi_{WX}(\zeta,R)=(\bp,\bx)$
where
\begin{equation}
\medbox{
\bp=\frac{1}{\sqrt{2}}\,\overline{\zeta'}\bsigma\zeta'
\qquad
\&
\qquad
\bx=\br
\quad
(t=0).
}
\label{px}
\end{equation}

We see that the energy $\cE=\vert\bp\vert$ is now given by
\begin{equation}
\vert\bp\vert=\frac{1}{\sqrt{2}}\,\Vert\zeta'\Vert^2
\label{cE}
\end{equation}
in spinorial terms.

\section{Geometric quantization of the model}

We propose now to quantize, following \cite{SSD}, the above classical model using the technique of geometric quantization.

\subsection{Prequantization}\label{prequantization}

Let us recall that a \textbf{prequantum manifold} above a symplectic manifold $(X,\omega)$ is a principal circle-bundle $\pi_{YX}:Y\to{}X$ with connection $\alpha/\hbar$ whose curvature descends to $X$ as $\omega/\hbar$, i.e., such that $d\alpha=\pi_{YX}^*\omega$ \cite{SSD}.\footnote{\label{zY}We will denote by $z\to{}z_Y$ the $\rU(1)$-action on $Y$.} See also \cite{Kos,SW,Woo} for an equivalent definition in terms of line-bundles.\footnote{Alternatively, prequantization can be understood in terms of the inequivalent forms of the ``Dirac-Feynman factor'' $\exp(iS/\hbar)$, where $S$ is the classical action \cite{Hor,Hor2}.
}

\goodbreak

\medskip$\bullet$
This (not so well-known) intermediate geometric structure, $(Y,\alpha)$, in-between the Poincar\'e group, $(W^{10},\varpi)$, and the space of classical  motions, $(X^6,\omega)$, will actually prove crucial in the derivation of the \textbf{massless Dirac equations} describing the quantum chiral particles in Minkowski spacetime. This remark applies, of course, to other prequantizable homogeneous symplectic manifold physically relevant to describe elementary particles.

\medskip$\bullet$
Let us rewrite, for convenience, the $1$-form (\ref{varpi}) of $W^{10}$ in the new guise
\begin{equation}
\medbox{
\varpi=R\cdot{}dP+\chi\hbar\frac{\barZ{}dZ}{i}
\qquad
\hbox{where}
\qquad
Z=e^{-\frac{i}{\hbar}P\cdot{}R}\zeta.
}
\label{varpibis}
\end{equation}

\goodbreak

We now claim that $Y^7=W^{10}/(\ker\varpi\cap\ker{}d\varpi)$ is the sought prequantum bundle over $(X^6,\omega)$, endowed with the \textbf{prequantum $1$-form} $\alpha$ such that $\varpi=\pi_{WY}^*\alpha$.\footnote{\label{Y7}Let us mention that the characteristic distribution $\ker\varpi\cap\ker{}d\varpi$ of $(W^{10},\varpi)$ is given by (\ref{kersigma}) with $\lambda=0$. It is therefore $3$-dimensional; its foliation corresponds to Wigner-Souriau trans\-lations~\cite{DHchiral} along $P^\perp$, so that $Y^7\cong{}W^{10}/\bbR^3$. By construction, the $1$-form $\varpi$ of $W^{10}$ descends to $Y^7$ as the $1$-form~$\alpha$ given by (\ref{alpha}). N.B. It is easily verified that $\zeta'$ (as well as $Z'$!) is a first-integral of the characteristic distribution of $\varpi$; it hence passes to $Y^7$. 
}
As in \cite{SSD}, the prequantum bundle, $Y^7$, is readily identified with the section of $W^{10}$ described by the pairs $(Z',\bx)$ where 
\begin{equation}
Z''=\frac{Z'}{\Vert{}Z'\Vert^2}
\qquad
\&
\qquad
\bx=\br 
\qquad(t=0)
\label{Z''x}
\end{equation}
with $Z'\in\bbC^2\setminus\{0\}$, and $\br\in\bbR^3$.
In view of (\ref{varpibis}) and (\ref{Z''x}), this immediately yields
\begin{equation}
\medbox{
\alpha=-\bx\cdot{}d\bp-\chi\hbar\frac{d\vert\bp\vert}{2i\vert\bp\vert}+\chi\hbar\frac{\overline{Z''}dZ'}{i}
}
\label{alpha}
\end{equation}
where $\bp\in\bbR^3\setminus\{0\}$ (see (\ref{px})) reads now
\begin{equation}
\medbox{
\bp=\frac{1}{\sqrt{2}}\,\overline{Z'}\bsigma{}Z'.
}
\label{p}
\end{equation}

\goodbreak

The $1$-form $\alpha/\hbar$ is clearly a $\rU(1)$-connection form; indeed $d\alpha=\pi_{YX}^*\omega$ where~$\omega$ is as in (\ref{omega}), and $\alpha(\delta(Z',\bx))=\hbar$ where $\delta(Z',\bx)=(i\chi{}Z',0)$ is the fundamental vector field of the 
$\rU(1)$-action on $Y^7$.
The projection $\pi_{YX}:\bbC^2\setminus\{0\}\times\bbR^3\to\bbR^3\setminus\{0\}\times\bbR^3$ is therefore given by $\pi_{YX}(Z',\bx)=(\bp,\bx)$.

This is summarized by the following diagrams
$$
\begin{CD}
W^{10} @> \pi_{WY} >> Y^7
\strut\\
@V{\pi_{WV}}VV @VV{\pi_{YX}}V 
\strut\\
V^9 @> \pi_{VX} >> X^6
\strut\\
@V{\pi_{VE}}VV 
\strut\\
E^4  
\end{CD}
\qquad
\qquad
\qquad
\begin{CD}
(\zeta,R) @> \pi_{WY} >> (Z',\bx)
\strut\\
@V{\pi_{WV}}VV @VV{\pi_{YX}}V 
\strut\\
(P,Q,R) @> \pi_{VX} >> (\bp,\bx)
\strut\\
@V{\pi_{VE}}VV 
\strut\\
R  
\end{CD}
$$

\goodbreak

\subsection{Geometric quantization and the Weyl equation}\label{quantization}

As every quantization procedure, geometric quantization needs the intro\-duction of wave functions serving to provide on the one hand solutions of the sought wave equations, and, on the other hand, unitary irreducible representations of the classical symmetry group at hand, e.g., the Poincar\'e group.

\goodbreak

\medskip$\bullet$
Geometric quantization primarily views \textbf{wave functions} as smooth $\rU(1)$-equivariant complex-valued functions, namely such that
\begin{equation}
\Psi:Y\to\bbC
\qquad
\&
\qquad
z_Y^*\Psi=z\Psi
\label{Psi}
\end{equation}
for all $z\in\rU(1)$; see Footnote \ref{zY}.\footnote{This definition legitimizes the overall phase factor affecting quantum wave functions as coming from the consideration of a circle bundle above the space of classical motions.}

\medskip$\bullet$
The next fundamental ingredient needed in geometric quantization is a \textbf{polarization}.\footnote{This is a geometrical object of symplectic geometry; no confusion with the optical notion of polarization of light!} Here, we will merely consider a real polarization of a symplectic manifold $(X^{2n},\omega)$, i.e., a (fiberwise) $n$-dimensional isotropic subbundle $F\subset{}TX$, i.e., verifying $\omega(F,F)=0$. 

This choice generalizes that of the ``position'' or ``momentum'' representation in quantum mechanics. The \textbf{$F$-polarized wave functions} of $(Y,\alpha)$ over $(X,\omega)$ are the $F^\sharp$-constant wave function $(\ref{Psi})$, viz.,
\begin{equation}
F^\sharp\Psi=0
\label{FsharpPsi=0}
\end{equation}
where $F^\sharp$ is the horizontal lift of $F$ to $(Y,\alpha)$; see also \cite{Kos,SW,Woo}. 

\medskip$\bullet$
Returning to our model $(X^6,\omega)$, we choose the most natural polarization, i.e., the \textbf{vertical polarization}, $F$, whose leaves, $\bp=\const.$, are the fibres of the bundle $X^6\to\cC^+$; see (\ref{X6}).\footnote{In view of~(\ref{omega}), we clearly have $\omega(F,F)=0$.} It turns out that $F=\ker\phi$, where
\begin{equation}
\phi=\frac{dp_1\wedge{}dp_2\wedge{}dp_3}{\vert\bp\vert}
\label{phi}
\end{equation}
is the \textbf{polarizer} \cite{Duv2,DE} defined by the (pull-back of the) canonical Poincar\'e-invariant $3$-form of the punctured future light-cone $\cC^+$. Note that $F$ is indeed a polarization iff
$$
\omega\wedge\phi=0
\qquad
\&
\qquad
\rank\,\phi=3.
$$
An equivalent approach has been devised in \cite{SW,Woo} to geometrically quantize, e.g., relativistic massless particle models.

\medskip$\bullet$
We are ready to express the $F$-polarized wave functions of the model using this formalism.
It has been shown \cite{DE} in full generality that those are the solutions $\Psi\in{}C^\infty(Y,\bbC)$ of the differential equation  
\begin{equation}
\phi\wedge{}D\Psi=0
\qquad
\hbox{where}
\qquad
D\Psi=d\Psi-\frac{i}{\hbar}\alpha\Psi
\label{phiDPsi=0}
\end{equation}
is the covariant derivative of the wave function, $\Psi$, of $(Y,\alpha)$. These wave functions, $\Psi$, are thus ``covariantly constant'' along the polarizer, $\phi$.


\medskip$\bullet$
Let us first examine the case, $\chi=-1$, of negative helicity in the above  model. 
We find, with the help of (\ref{alpha}) and (\ref{phi}), that the solutions, $\Psi$, of Equation (\ref{phiDPsi=0}) must satisfy\footnote{Easy calculation leads to $\overline{Z}dZ
=-d\overline{Z''}Z'-d\vert\bp\vert/(2\vert\bp\vert)$.}
\begin{eqnarray*}
dp_1\wedge{}dp_2\wedge{}dp_3
&
\wedge
&
\displaystyle
\Big[\,\frac{\partial\Psi}{\partial\bx}d\bx+\frac{\partial\Psi}{\partial\bp}d\bp+\frac{\partial\Psi}{\partial Z'}dZ'+\frac{\partial\Psi}{\partial \overline{Z''}}d\overline{Z''}\\[6pt]
&&
\displaystyle
+\left(\frac{i}{\hbar}\bx\cdot{}d\bp-\frac{d\vert\bp\vert}{2\vert\bp\vert}-\overline{dZ''}Z'\right)\Psi
\Big]=0
\end{eqnarray*}
which gives $\partial\Psi/\partial\bx=0$, $\partial\Psi/\partial{}Z'=0$, and $\partial\Psi/\partial\overline{Z''}=Z'\Psi$. This entails that $\overline{Z''}\,\partial\Psi/\partial\overline{Z''}=\overline{Z''}Z'\Psi=\Psi$; hence $\Psi$ is homogeneous of degree $1$ in $\overline{Z''}$, viz.,
$
\Psi(Z',\bx)=\overline{Z''}\psi'(P)
$
for some $\psi'\in{}C^\infty(\cC^+,\bbC^2)$. It thus follows that 
\begin{equation}
\Psi(Z',\bx)=\barZ\,\psi(P)
\qquad
\hbox{where}
\qquad
\psi(P)=\left(
\begin{array}{c}
2Z'\\
0
\end{array}
\right)\Psi(Z',\bx).
\label{Psichi=-1}
\end{equation}
Using (\ref{gammas}), (\ref{cE}), and (\ref{p}), one furthermore shows that\footnote{Indeed, if $\xi=Z'/\Vert{}Z'\Vert\in{}S^3$, one has $\sigma(\bu)=2\xi\overline{\xi}-1$.}
$$
\gamma(P)=\frac{1}{\sqrt{2}}
\left(
\begin{array}{cc}
0&Z'\overline{Z'}\\
\Vert{}Z'\Vert^2-Z'\overline{Z'}&0
\end{array}
\right).
$$


We have just proven, if $\chi=-1$, that the \textbf{space of so-polarized wave functions} (\ref{Psichi=-1}) is isomorphic to the space of spinors $\psi\in{}C^\infty(\cC^+,\bbC^{2,2})$ satisfying
\begin{equation}
\medbox{
\gamma(P)\psi(P)=0
\qquad
\&
\qquad
\gamma_5\psi(P)=i\chi\psi(P).
}
\label{chiPolarizedWavefunctions}
\end{equation}

\medskip$\bullet$
The fundamental equations in (\ref{chiPolarizedWavefunctions}) hold for $\chi=\pm1$ \cite{SSD}.

\goodbreak

\medskip$\bullet$
To finish, let us introduce the partial Fourier transform
\begin{equation}
\tpsi(R)=\int_{
\cC^+}{\!\!\psi(P)\,e^{\frac{i}{\hbar}P\cdot{}R}\,\phi
}
\label{tildepsi}
\end{equation}
where $\psi$ is now supposed to be compactly supported on the forward null light-cone, $\cC^+$. 

Routine computation shows, with recourse to (\ref{chiPolarizedWavefunctions}) and (\ref{tildepsi}), that $\tpsi\in{}C^\infty(E^4,\bbC^{2,2})$ is \textit{indeed} the general solution of the \textbf{Weyl equation} of helicity $\chi=\pm1$, namely
\begin{equation}
\medbox{
\gamma^\mu\partial_\mu\tpsi=0
\qquad
\&
\qquad
\gamma_5\tpsi=i\chi\tpsi.
}
\label{WeylEq}
\end{equation}


The genesis of the Weyl equations by geometric quantization is originally due to Souriau \cite{SSD}. It has later been revisited by Simms \& Woodhouse \cite{SW,Woo} using the spinorial/twistorial formalism of Penrose; see also \cite{DE}.

\medskip$\bullet$
The universal covering, $G$, of the Poincar\'e group consists of the matrices 
$$
g=
\left(
\begin{array}{cc}
\Ad(A)&\gamma(C)\\
0&1
\end{array}
\right)
\qquad
\hbox{where}
\qquad
A\in\Spin(3,1)\quad \& \quad C\in\bbR^{3,1}.
$$ 
Since $G\cong{}W^{10}$, its left-action reads $g\cdot(\zeta,R)=(A\zeta,\varrho(A)R+C)$. The latter induces, via pull-back, a natural (right-)action on the wave function~(\ref{Psichi=-1}) so that $g\cdot\psi(P)=A^{-1}\psi(\varrho(A)P)$. Taking into account the Lorentz-invariance of the polarizer (\ref{phi}) in the integral (\ref{tildepsi}), immediately leads to the following \textbf{(anti-)representation} of the universal covering of the Poincar\'e group on the solutions (\ref{tildepsi}) of the Weyl equation (\ref{WeylEq}), viz.,
\begin{equation}
g\cdot\tpsi(R)=A^{-1}\tpsi(\varrho(A)R+C)
\label{G-action}
\end{equation}
for all $g=(A,C)\in\Spin(3,1)\ltimes\bbR^{3,1}$.


\medskip$\bullet$
Let us express our Weyl system (\ref{WeylEq}) in terms of $2$-component spinors, $\tpsi_\chi\in{}C^\infty(E^4,\bbC^2)$ corresponding to helicity $\chi=\pm1$. Using the representation~(\ref{gammas}) of the gamma matrices, those show up as follows
$$
\tpsi
=\left(
\begin{array}{c}
\tpsi_-\\[4pt]
\tpsi_+
\end{array}
\right).
$$

\goodbreak

The Dirac operator, $\sfD=\gamma^\mu\partial_\mu$, then reads
$$
\sfD=
\left(
\begin{array}{cc}
0&-\sigma^j\partial_j+\partial_t\\
\sigma^j\partial_j+\partial_t&0
\end{array}
\right).
$$
so that $\sfD\tpsi=0$ in (\ref{WeylEq}) yields the \textbf{equivalent form of the Weyl equation}, namely
\begin{equation}
\sigma^j\partial_j\tpsi_\chi=\chi\,\partial_t\tpsi_\chi
\label{WeylEqBis}
\end{equation}
with $\partial_j=\partial/{\partial r^j}$ for all $j=1,2,3$, where $\br=(r^1,r^2,r^3)$ stand for the \textbf{position co\-ordinates}, i.e., the \textit{bona fide} spatial translation coordinates of the Poincar\'e group.

\bigskip

\textbf{Acknowledgements:} It is a pleasure to thank P. Horv\'athy for his help and encouragement.





\begin{thebibliography}{99}
       
\bibitem{Duv1}
 C.~Duval,
 ``On the prequantum description of spinning particles in an external gauge field'',
{\em in Proc. Coll. Int. CNRS ``Differential Geometrical Methods in
Mathematical Physics''},
Aix en Provence-Salamanca 1979,
(P.L.~Garcia, A.~P\'erez-Rendon, J.-M.~Souriau Eds),
pp 49--66,
Lecture Notes in Mathematics \textbf{836},
Springer-Verlag (1980).

\bibitem{Duv2}
 C.~Duval,
 ``On the polarizers of compact semi-simple Lie groups. Applications'',
{\em Ann. Inst. \allowbreak H.~Poincar\'e}
{\bf A 34}
(1981)
95--115.

\bibitem{DE}
 C.~Duval and J.~Elhadad,
 ``Geometric Quantization and Localization of Relativistic Spin Systems'',
{\em in Proc. AMS of the 1991 ``Joint Summer Research
Conference on Mathematical Aspects of Classical Field Theory''},
Seattle 1991,
(M.J.~Gotay, J.E.~Marsden, V.E.~Moncrief Eds),
Contemporary  Mathematics {\bf 132} (1992), 317--330.

\bibitem{DHchiral}
  C.~Duval and P.~A.~Horvathy,
``Chiral fermions as classical massless spinning particles,''
{\em Phys. Rev.} {\bf D91} 045013 (2015).
            
\bibitem{DEHZ}
  C.~Duval, M.~Elbistan, P.~A.~Horvathy and P.-M.~Zhang,
 ``Wigner-Souriau translations and Lorentz symmetry of chiral fermions,''
{\em Phys. Lett.} {\bf B 742} (2015) 322 
  
 
\bibitem{Hor}
P. A. Horv\'athy,
``Classical action, the Wu-Yang phase factor and prequantization,''	
Proc. Int. Coll. on {\it Diff. Geom. Meths. in Math. Phys}., Aix-en Provence '79. Ed. Souriau. Springer Lecture Notes in Mathematics {\bf 836}, 67 (1980).

\bibitem{Hor2}
 P.~A.~Horv\'athy,
``Prequantization From Path Integral Viewpoint,''
  Lecture Notes in Mathematics\  {\bf 905} (1982) 197.

\bibitem{Kos}
B. Kostant,
``On certain unitary representations which arise from a quantization theory,''
in {\em Group representations in mathematics and physics}
(V. Bargmann, Ed.),
pp. 237--253,
Proceedings Battelle Seattle 1969 Rencontres; LNP \textbf{6},
Springer-Verlag,
Berlin,
1970.

\bibitem{SW}
D. J. Simms and N. M. J. Woodhouse,
\textsl{Lectures on Geometric Quantization},
Lecture Notes in Physics \textbf{53}, Springer-Verlag Berlin-Heidelberg-New York, 1976.

\bibitem{SSD}
J.-M.~Souriau,
\textsl{Structure des syst\`emes dynamiques}, Dunod (1970), \copyright1969). 
\textsl{Structure of Dynamical Systems. A Symplectic View of Physics}, Birkh\"auser, Boston (1997). 

\bibitem{Woo}
N. M. J. Woodhouse,
\textsl{Geometric quantization}. 
Second edition,
Oxford University Press, New York, 1992.

\end{thebibliography}
\end{document}